\documentclass[preprint,showpacs,preprintnumbers,amsmath,amssymb]{revtex4}

\usepackage{graphicx,color}

\begin{document}

\title{
Constraining 
extra gravitational wave polarizations 
with Advanced LIGO, Advanced Virgo and KAGRA and 
upper bounds from GW170817
} 
\author{Yuki Hagihara${}^{1}$}
\author{Naoya Era${}^{1}$}
\author{Daisuke Iikawa${}^{1}$}
\author{Atsushi Nishizawa${}^{2,3}$}
\author{Hideki Asada${}^{1}$} 
\affiliation{
${}^{1}$
Graduate School of Science and Technology, Hirosaki University,
Aomori 036-8561, Japan\\
${}^{2}$
Research Center for the Early Universe, 
University of Tokyo,
Tokyo 113-0033, Japan\\
${}^{3}$
Kobayashi-Maskawa Institute for the Origin of Particles and the Universe, 
Nagoya University, Nagoya 464-8602, Japan 
}

\date{\today}

\begin{abstract} 
General metric theories of gravity in four-dimensional spacetimes 
can contain at most six polarization modes (two spin-0, two spin-1 and two spin-2) 
of gravitational waves (GWs). 
It has been recently shown that, with using four GW non-coaligned detectors, 
a direct test of the spin-1 modes can be done in principle 
separately from the spin-0 and spin-2 modes 
for a GW source in particular sky positions 
[Hagihara et al., Phys. Rev. D 98, 064035 (2018)]. 
They have found particular sky positions 
that satisfy a condition of killing completely the spin-0 modes in 
a so-called null stream which is a linear combination of 
the signal outputs to kill the spin-2 modes. 
The present paper expands the method to 
discuss a possibility 
that the spin-0 modes are not completely but effectively suppressed 
in the null streams to test the spin-1 modes separately from the other modes, 
especially with an expected network of Advanced LIGO, 
Advanced Virgo and KAGRA. 
We study also a possibility that the spin-1 modes are substantially suppressed 
in the null streams to  test the spin-0 modes separately from the other modes, 
though the spin-1 modes for any sky position cannot be completely killed in the null streams. 
Moreover, we find that the coefficient of the spin-0 modes in the null stream 
is significantly small for the GW170817 event, 
so that an upper bound can be placed 
on the amplitude of the spin-1 modes as $< 6 \times 10^{-23}$. 
\end{abstract}

\pacs{04.80.Cc, 04.80.Nn, 04.30.-w}

\maketitle

\section{Introduction}
Gravitational waves (GWs) in Einstein's theory of general relativity (GR), 
spacetime ripples that propagate at the speed of light, 
can be formulated in terms of the so-called ``tensor" modes 
which are spin-2 transverse and traceless (TT) parts of the spacetime metric 
\cite{Einstein1916, Einstein1918}. 
The speed of the GW propagation has been confirmed surprisingly at the level of 
$\sim O(10^{-15})$ 
by the GW170817 event with an electromagnetic (EM) counterpart 
as the first observation of GWs from a NS-NS merger 
\cite{GW170817}. 
In the rest of the present paper, we assume that the speed of GWs 
is almost the same as the speed of light. 
For a GW source with an EM counterpart, 
we know its precise location on the sky. 
In the present paper, we use essentially 
the information mainly on the direction 
of the GW source. 
Note that the assumption on the GW speed is not crucial 
in the present method. 
Extra GW modes may be delayed. For a few seconds (or minutes) delay, 
we will be able to still recognize that the delayed signals come from 
the same GW source and thus the sky location of the GW source is known from 
EM observations. 

Alternative theories to GR, more specifically  
general metric theories of gravity in four-dimensional spacetimes 
can predict extra degrees of freedom 
with spin 0 and spin 1, 
which are usually called scalar and vector modes, respectively 
\cite{Eardley, PW}. 
Future detection of scalar or vector GW polarization would provide 
serious evidence against GR. 
Or future GW polarization tests would place a strong constraint on 
scalar and vector modes of GWs, which will lead to 
a new test of modified gravity theories, 
some of which might be ruled out. 
Therefore, many attempts for GW polarization tests 
by using not only bursts, pulsars, compact binary coalescences, 
but also stochastic sources have been discussed 
(e.g. \cite{Isi2015, Isi2017, Svidzinsky, pulsar, Nishizawa2009, Hayama, Nishizawa2018}).

The GW astronomy has just started and 
aLIGO-Hanford (H), aLIGO-Livingston (L) and Virgo (V) 
have already detected GW signals. 
HLV will start the O3 observation run in 
April this year. 
However, three detectors are not enough 
for distinguishing every polarization of GWs. 
The construction of another kilometer-scale interferometer called KAGRA (K) 
is currently urged to join the GW detector network as a fourth detector 
by the end of the O3 run. 
See Reference \cite{LVK} for a comprehensive review on 
the expected network of Advanced LIGO, Advanced Virgo and KAGRA 
(denoted as HLVK). 
Therefore, it is currently interesting to examine how to probe extra 
GW polarizations by HLVK.

It was thought that five or more non-coaligned GW interferometers were needed 
to directly test the extra degrees of freedom of GW with spin 0 and 1 
separately from each other. 
In fact, Chatziioannou, Yunes and Cornish have argued null streams 
for six (or more) GW detectors to probe GW polarizations 
\cite{CYC}. 
The null stream approach was introduced first 
by G\"ursel and Tinto \cite{GT} 
and was extended by Wen and Schutz \cite{WS} 
and Chatterji et al. \cite{null-papers}, 
where the idea behind the null stream method is that there exists 
a linear combination of the data from a network of detectors, 
such that the linear combination called a null stream can contain 
no tensor modes but only noise in GR cases. 
G\"ursel and Tinto \cite{GT} proposed the use of the null stream 
in order to understand the noise behavior. 

Assuming that, for a given source of GWs, 
its sky position is known, as is the case of GW events with 
an electromagnetic (EM) counterpart such as GW170817, 
however, 
it has been recently found that there are particular sky positions 
that satisfy a condition for the spin-0 modes to be killed completely 
in the null streams in which the spin-2 GW modes disappear 
and thus the spin-1 modes can be directly 
tested separately from the spin-0 and spin-2 modes, 
even with using only four GW non-coaligned detectors, 
though the strain output at a detector may contain the spin-0 modes 
\cite{Hagihara}. 
They have found that there are seventy sky positions exactly at which 
the spin-0 modes are killed in the null streams. 

However, the area of the seventy points is zero. 
Does this mean that the probability of such a potentially important event 
is negligible? 
In order to answer this, 
we shall examine whether a GW source near one of the seventy sky positions 
can be used for testing (e.g. placing an upper bound on) spin-1 GW modes, 
if the spin-0 modes are substantially suppressed 
even though they are not perfectly killed. 

How large is the sky area where the spin-0 modes are significantly suppressed 
and thus the spin-1 modes can be tested separately from the other modes? 
This is an interesting subject, especially for the near future GW observations 
with using the HLVK network. 

The main purpose of this paper is to examine 
how large is the probability that spin-0 modes 
are substantially suppressed and thus spin-1 GW modes can be tested separately 
from the other GW polarizations by HLVK. 
This paper is organized as follows. 
In Section II, we describe null streams 
for four non-coaligned detectors.  
In Section III, we discuss a possibility 
that the spin-0 modes are substantially suppressed in the null streams 
and hence the spin-1 modes can be tested within the noise level, 
especially with an expected network of HLVK. 
In Section IV, we study also a possibility that the spin-1 modes are 
almost killed in the null streams and thus the spin-0 modes 
become testable separately 
from the other modes. 
In Section V, we discuss a constraint on the spin-1 modes 
by the GW170817 event. 
Section VI is devoted to a summary of this paper. 
Throughout this paper, 
Latin indices $a, b, \cdots$ run from 1 to 4 
corresponding to four detectors.

\section{Null streams for four non-coaligned detectors}
Let us assume that there exist four non-coaligned 
detectors with uncorrelated noise. 
Each detector is labeled by $a$ ($a=1,2,3$ and $4$). 
We assume also that, for a given GW source, we know its sky position, 
as is the case of GW events with an EM counterpart such as GW170817. 
By the second assumption, 
we know exactly how to shift the arrival time of the GW 
from detector to detector. 

For each detector, 
the 
signal from a GW source 
at the sky location denoted as $(\theta, \phi)$ is written as 
\begin{align}
S_a =& 
F_a^{+} h_{+} + F_a^{\times} h_{\times} 
\nonumber\\
&+ F_a^{S} h_S + F_a^{L} h_L 
\nonumber\\
&+ F_a^{V} h_V + F_a^{W} h_W + n_a ,
\label{S0}
\end{align}
where 
$h_{+}$ and $h_{\times}$ denote the spin-2 modes called 
the plus and cross mode, respectively, 
$h_S$ and $h_L$ denote the  
spin-0 
modes called 
the breathing and longitudinal mode, respectively, 
and 
$h_V$ and $h_W$ denote the spin-1 modes 
often called the vector-$x$ and vector-$y$ mode, respectively 
\cite{footnote0}, 
$F_a^{+}$, $F_a^{\times}$, $F_a^{S}$, $F_a^{L}$, 
$F_a^{V}$ and $F_a^{W}$ 
are the antenna patterns for GW polarizations 
\cite{PW,Nishizawa2009,ST} 
and 
$n_a$ denotes noise at the detector. 
The antenna patterns are functions of 
a GW source location  
$\theta$ and $\phi$ \cite{footnote1}. 
In our numerical calculations, 
$\theta$ and $\phi$ denote the 
latitude and longitude, respectively. 

By noting $F_a^{S} = - F_a^{L}$ 
\cite{footnote2}
that was shown by 
Nishizawa et al. in \cite{Nishizawa2009}, 
Eq. (\ref{S0}) can be simplified as 
\begin{align}
S_a =& 
F_a^+ h_{+} + F_a^{\times} h_{\times} 
\nonumber\\
&+ F_a^S (h_S - h_L) 
\nonumber\\
&+ F_a^V h_V + F_a^W h_W + n_a . 
\label{S}
\end{align}
Note that the effects of $h_S$ on the detector 
are exactly the same with the opposite sign as those of $h_L$. 
Hence, GW interferometers can test 
only the difference as $h_S - h_L$, 
one combined spin-0 mode. 

Eq. (\ref{S}) may suggest that five or more non-coaligned detectors 
are needed for measuring five components $h_{+}$, $h_{\times}$, 
$h_V$, $h_W$ and $h_S - h_L$. 
This is related with the inversion of a $5 \times 5$ matrix. 
Namely, the existence of the inverse matrix is assumed implicitly 
when the above suggestion holds. 
In fact, Hagihara et al. have found that there are exceptional cases, 
for which spin-0 modes are completely killed and 
thus spin-1 modes can be tested separately from 
spin-2 and spin-0 modes. 
Their study is entirely based on the null streams. 
The null stream was introduced first by G\"ursel and Tinto \cite{GT}, 
who considered only the purely tensorial modes $h_{\times}$ and $h_{+}$. 
Let us imagine, for its simplicity, an ideal case that 
noise is negligible in Eq. (\ref{S}). 
Then, by straightforward calculations, one can obtain 
a null stream \cite{GT} as, for three detectors as $a=1, 2$ and $3$ for instance, 
\begin{align}
\delta_{23}S_1+\delta_{31}S_2+\delta_{12}S_3 = 0 ,  
\label{Null-123}
\end{align}
where we define 
\begin{align}
\delta_{ab} \equiv F_a^+ F_b^{\times} - F_b^+ F_a^{\times} . 
\end{align}
Namely, signal outputs at the three detectors must satisfy 
Eq. (\ref{Null-123}), 
provided that the signals are made only 
from the spin-2 waves. 

For four GW detectors with noise, the null streams become 
\begin{align}
\delta_{23}S_1+\delta_{31}S_2+\delta_{12}S_3 
&= 
\delta_{23}n_1+\delta_{31}n_2+\delta_{12}n_3 , 
\label{Null-123-n}
\\
\delta_{34}S_2+\delta_{42}S_3+\delta_{23}S_4 
&= 
\delta_{34}n_2+\delta_{42}n_3+\delta_{23}n_4 , 
\label{Null-234-n}
\\
\delta_{41}S_3+\delta_{13}S_4+\delta_{34}S_1 
&= 
\delta_{41}n_3+\delta_{13}n_4+\delta_{34}n_1 , 
\label{Null-341-n}
\\
\delta_{12}S_4+\delta_{24}S_1+\delta_{41}S_2 
&= 
\delta_{12}n_4+\delta_{24}n_1+\delta_{41}n_2  . 
\label{Null-412-n}
\end{align}
The number of independent equations 
in Eqs. (\ref{Null-123-n})-(\ref{Null-412-n}) is two 
\cite{Hagihara}.

Next, we incorporate scalar and vector polarization modes. 
Only the tensor modes cancel out in the tensor null stream 
as Eq. (\ref{S}). 
The scalar and vector modes can exist in the null stream. 
Two null streams can be written as 
\cite{Hagihara} 
\begin{align}
P_a S_a &= (P_bF_b^S)(h_S - h_L) + (P_cF_c^V)h_V + (P_dF_d^W)h_W 
+ P_e n_e ,
\label{PS}
\\
Q_f S_f &= (Q_gF_g^S)(h_S - h_L) + (Q_hF_h^V)h_V + (Q_iF_i^W)h_W 
+ Q_j n_j , 
\label{QS}
\end{align}
where we use Eq. (\ref{S}) and 
the summation is taken over $a, \cdots, j=1, 2, 3$ and $4$. 
Note that the tensor null stream is built in and hence 
$h_{+}$ and $h_{\times}$ do not appear in the above equations. 
Without loss of generality, we can choose 
$P_a$ and $Q_a$ as 
$(P_a) = (\delta_{23}, \delta_{31}, \delta_{12}, 0)$ and 
$(Q_a) = (0, \delta_{34}, \delta_{42}, \delta_{23})$ 
for its simplicity, 
which are corresponding to Eqs. (\ref{Null-123-n}) and (\ref{Null-234-n}) 
in the previous paragraph. 

In the following sections, we shall examine Eqs. (\ref{PS}) and (\ref{QS}) 
in more detail. 
In numerical calculations for the HLVK network, 
we choose H=1, L=2, V=3 and K=4 
for $a=1, 2, 3$ and $4$ 
for its simplicity. 

\section{Suppression of spin-0 modes} 
Let us study the behavior of the spin-0 modes in the null streams 
by Eqs. (\ref{PS}) and (\ref{QS}), in which 
the coefficients of $h_S - h_L$ are $P_aF_a^S$ and $Q_aF_a^S$, respectively. 
Note that the tensor null stream is built in and 
hence $h_{+}$ and $h_{\times}$ do not exist in the equations. 

If $P_aF_a^S$ and $Q_aF_a^S$ are substantially small, 
spin-0 GW components can be taken to make a small contribution to the null streams. 
We look for quantitative criteria about whether or not the coefficients are small. 
In this paper, we do not assume a particular model of modified gravity. 
As a candidate for such criteria, therefore, 
we define 
a suppression factor by 
\begin{align}
\rho^S 
\equiv \frac{\max(|\mbox{PQF}^S|) - |\mbox{PQF}^S|}{\max(|\mbox{PQF}^S|)} , 
\label{suppression}
\end{align}
where $|\mbox{PQF}^S|$ denotes that larger one of $|P_aF_a^S|$ and $|Q_aF_a^S|$ 
and $\max(|\mbox{PQF}^S|)$ is the largest one of $|\mbox{PQF}^S|$ 
for the observed events. 
This approach is a possible extension of Reference \cite{Hagihara}, 
in which they focused only on the sky positions corresponding to 
$\rho^S =1$. 
In realistic situations, each detector has noise and thus 
$\rho^S = 1$ is too strict. 
Therefore, introducing the suppression factor in the discussion 
will make the present approach more practical for use. 

\begin{figure}
\includegraphics[width=16cm]{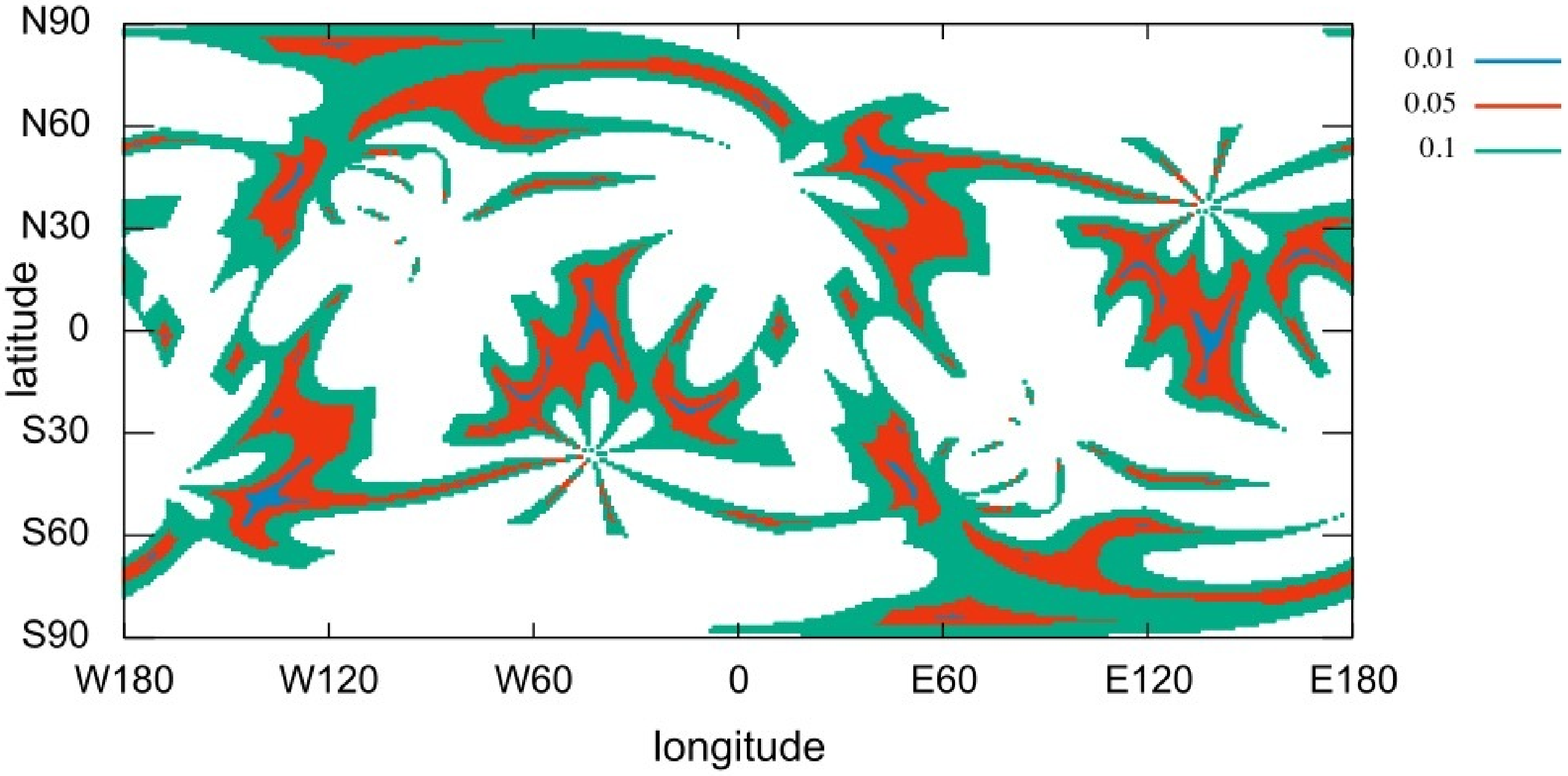}
\caption{
Contour map of 
${|\mbox{PQF}^S|}$ 
the larger one of the two scalar coefficients 
$|P_aF_a^S|$ and $|Q_aF_a^S|$ in the null streams 
for the HLVK network. 
The vertical and horizontal axes denote 
the latitude and longitude of a GW source, respectively. 
The maximum value of the scalar coefficient is 
around 0.5. 
}
\label{figure-contour-scalar}
\end{figure}

Figure  \ref{figure-contour-scalar} 
shows the sky map, which is the contour map for 
the coefficients of the scalar modes 
in the null streams by Eqs. (\ref{PS}) and (\ref{QS}). 
At some positions in the contour map, one can recognize an 
octapolar behavior of some curves. 
This behavior is because 
$\delta_{ab}$ is quadratic in $F_a^+$ and $F_b^{\times}$, where 
the GW antenna pattern functions are quadratic 
in trigonometric functions of $\theta$ and $\phi$. 
This figure shows that 
GW source positions with the suppression factor 
$\rho^S \geq 0.9$ 
have a significantly large area in the whole sky. 
We should note that 
domains with a much larger suppression factor 
$\rho^S \geq 0.99$ can be still recognized by eyes in the contour map. 
The probability for such largely suppressed events 
is not so negligible. 

\begin{figure}
\includegraphics[width=16cm]{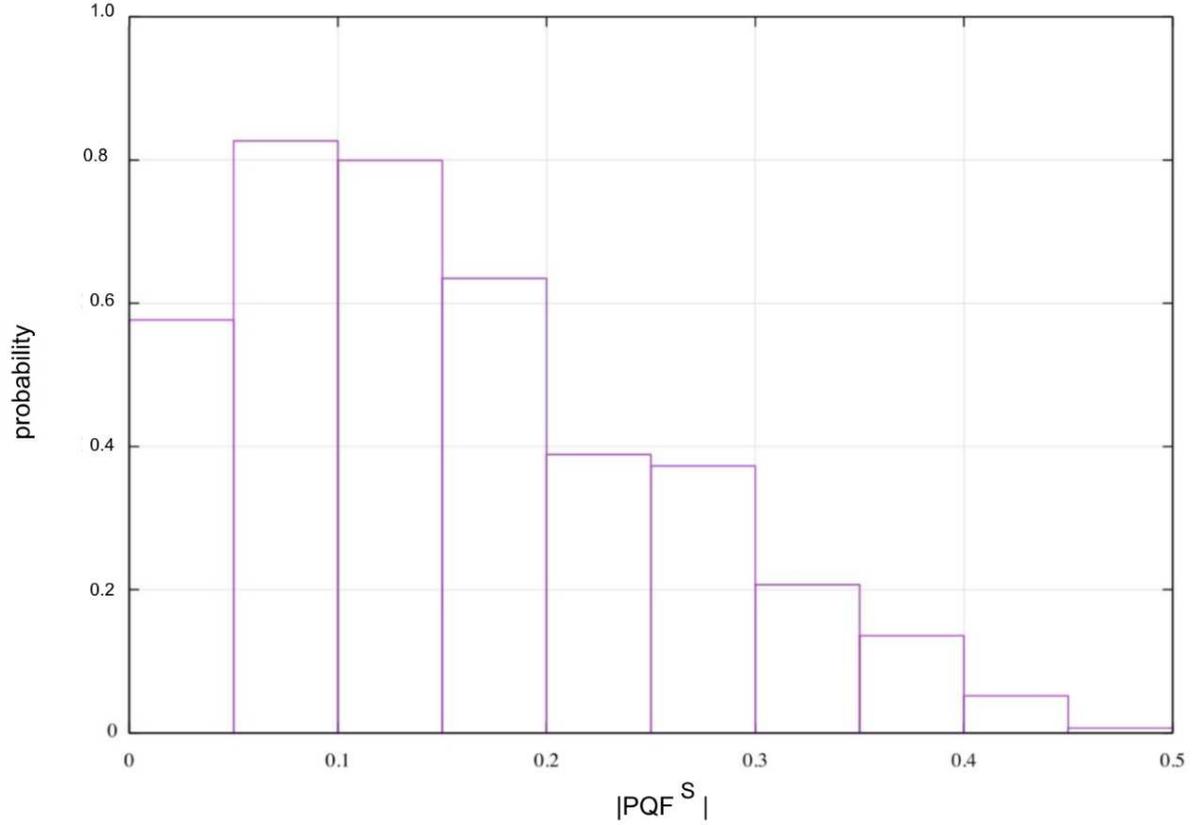}
\caption{
The vertical axis denotes the 
probability for each bin, 
where the horizontal axis denotes 
$|\mbox{PQF}^S|$, namely 
the larger one of $|P_aF_a^S|$ and $|Q_aF_a^S|$ 
in the null stream. 
We prepare numerically a random distribution of 10,000 events 
in the sky. 
The probability distribution has a single peak around $\sim 0.1$. 
The overall behavior of this histogram is confirmed numerically 
by changing the total number of events such as 20,000 and 100,000. 
}
\label{figure-histogram-scalar}
\end{figure}

\begin{figure}
\includegraphics[width=16cm]{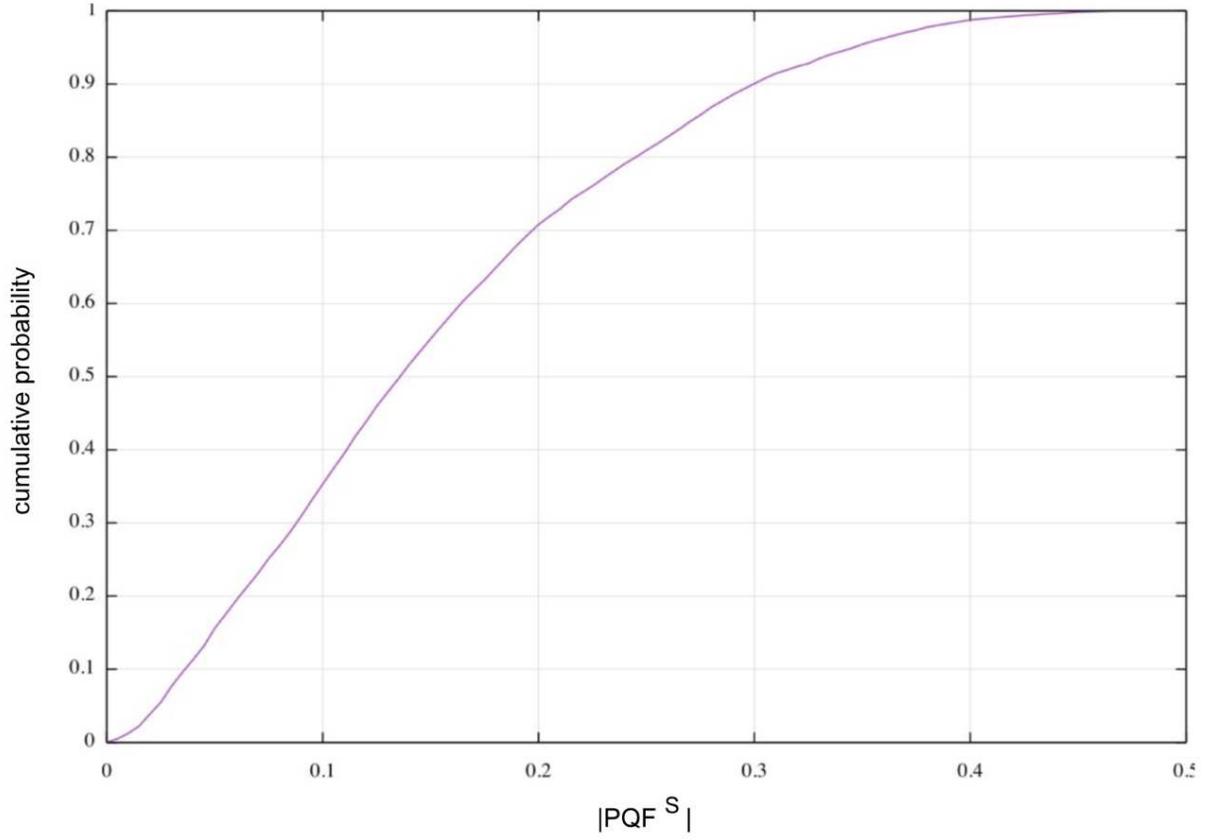}
\caption{
Event rate v.s. the threshold on $|\mbox{PQF}^S|$, 
the larger one of $|P_aF_a^S|$ and $|Q_aF_a^S|$ 
for the scalar mode, 
corresponding to Figure \ref{figure-histogram-scalar}. 
The cumulative 
probability in the vertical axis increases 
as the threshold for the suppression factor becomes larger. 
}
\label{figure-accumulate-scalar}
\end{figure}

We discuss the probability distribution of the coefficient 
$|\mbox{PQF}^S|$ 
in the null streams by numerically generating 10,000 GW events 
randomly located on the sky. 
Figure \ref{figure-histogram-scalar} 
shows the probability distribution 
as a function of the larger one between the two coefficients. 
The statistical fluctuation in each bin of the histogram is small 
($\sim$ a few percents), 
so that overall it cannot affect the shape of the histogram. 
This figure suggests that events with a 
large suppression factor may have a substantial probability. 
For this to be clearer, we plot the  
cumulative probability distribution of events. 
See figure \ref{figure-accumulate-scalar}. 
If we choose 
the threshold for the suppression factor 
on the scalar modes as  
$\rho^S \geq 0.8$ 
(corresponding to $|\mbox{PQF}^S| \leq 0.1$)  
for instance, then, nearly 30 percents of the total 
events can be used for a practical test 
of spin-1 polarizations. 
Even if we choose a tighter threshold as 
$\rho^S \geq 0.9$  
(corresponding to $|\mbox{PQF}^S| \leq 0.05$), 
nearly 20 percents of the events can be still used 
for a separate test of spin-1 modes. 
Roughly speaking, if five GW events with EM counterparts are detected 
in future, one of the events can be used for spin-1 tests. 

\begin{figure}
\includegraphics[width=16cm]{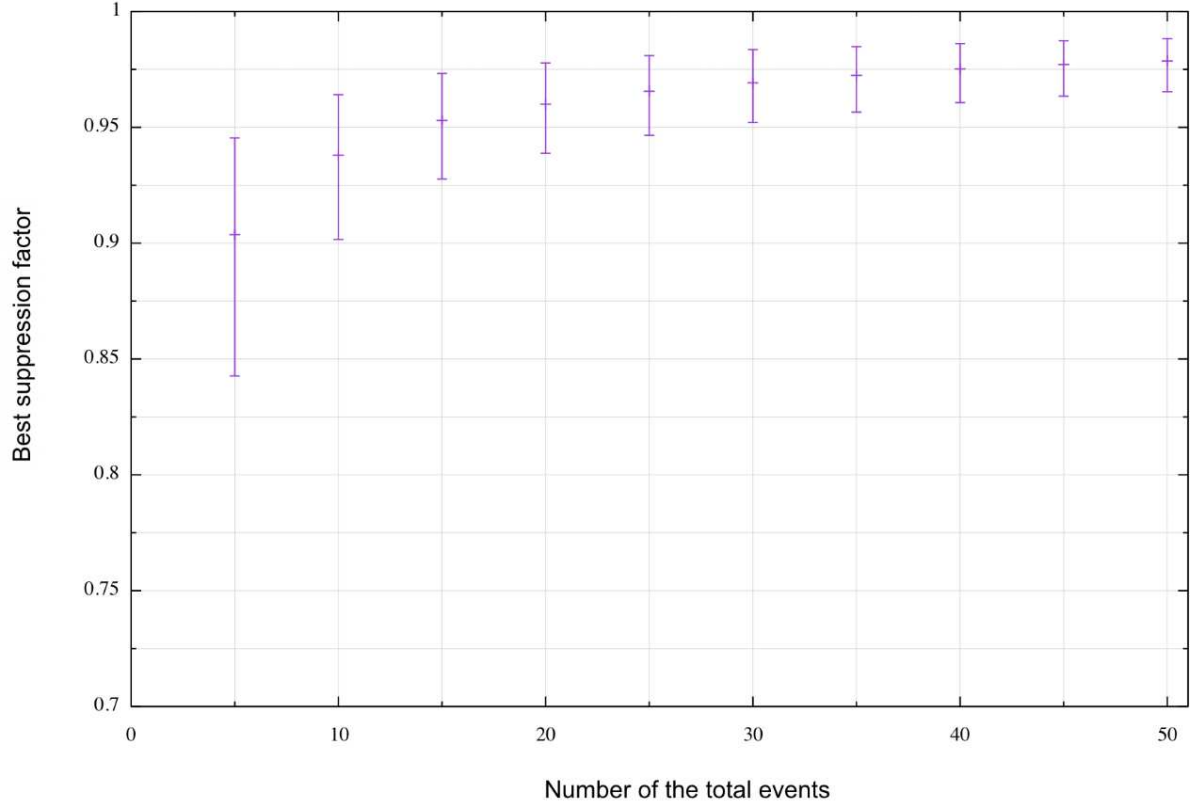}
\caption{
The vertical axis denotes the best suppression factor that is expected 
for a given total event number 
which is denoted as the horizontal axis. 
The error bars represent $1\sigma$ deviation. 
If the number of observed events is increased up to ten for example, 
the best suppression may reach 0.9 or more. 
For this case, a contribution of 
the scalar modes in the null streams is  
suppressed by a factor of ten or more and hence such an event 
can be practically used 
for a separate test of the spin-1 modes. 
}
\label{figure-eventnumber-scalar}
\end{figure}

Figure \ref{figure-eventnumber-scalar} shows the best suppression factor 
that can be expected for a given number of the total events. 
If about ten events are observed, the best suppression is likely to 
be 0.9 more or less, so that such an event with a large suppression 
can be used as a probe of the vector GW mode.

\section{Suppression of spin-1 modes} 
Next, we consider a suppression of the spin-1 modes in the null stream. 
This subject is new in the sense that Reference \cite{Hagihara} 
does not study the suppression of the spin-1 modes.  
First, we examine whether or not 
$P_aF_a^V$, $P_aF_a^W$, $Q_aF_a^V$ and $Q_aF_a^W$ vanish simultaneously 
at a sky position for the HLVK network. 
Our numerical result shows that they never do. 
This is a marked contrast to the spin-0 case, 
for which the two scalar coefficients $P_aF_a^S$ and $Q_aF_a^S$ can 
simultaneously vanish (and actually do at seventy positions in the sky 
\cite{Hagihara}). 
This can be understood by noting that 
two curves on a surface can intersect at some point, 
if they are not lines parallel to each other.  
On the other hand, 
three (or more) curves do not generally 
pass through the same point except for very special cases.

For the practical purpose, however, if all of 
$P_aF_a^V$, $P_aF_a^W$, $Q_aF_a^V$ and $Q_aF_a^W$ are small 
in the neighborhood of 
a certain sky location, 
a contribution of the spin-1 modes to 
the right-hand sides of the null streams is 
so small that  
such a case can give us an opportunity 
for a test of spin-0 polarizations. 
Therefore, let us examine how often the simultaneous suppression 
of the four vector coefficients 
in Eqs. (\ref{PS}) and (\ref{QS}) 
can occur. 

First, we should note that $P_aF_a^V$ and $P_aF_a^W$ are related with 
each other, because of the spin-1 nature. 
One can show that $(P_aF_a^V)^2 + (P_aF_a^W)^2$ is invariant 
for the spatial rotation around the axis of the GW propagation direction 
as follows.

Let us consider a rotation around the GW propagation axis 
with the rotation angle denoted as $\psi$. 
The GW spin-1 modes are transformed as 
\begin{align}
\left(
\begin{array}{c}
h_V^{\prime}\\
h_W^{\prime}
\end{array}
\right)
&= 
\left(
\begin{array}{cc}
\cos\psi & -\sin\psi\\
\sin\psi & \cos\psi
\end{array}
\right)  
\left(
\begin{array}{c}
h_V \\
h_W 
\end{array}
\right) , 
\label{vector}
\end{align}
where the prime denotes the rotation with $\psi$. 
In Eq. (\ref{PS}), the left-hand side including the detectors' signals 
do nothing to do with the above rotation with $\psi$. 
The first term and the last one of the right-hand side in Eq. (\ref{PS})  
are spin-0 and the detectors' noise, respectively,  
and thus they are independent of the $\psi$ rotation. 
As a result, the sum of the remaining (second and third) terms 
of the right-hand side 
is independent of the $\psi$ rotation, 
though each term may change by the rotation. 
Namely, we find 
for any $h_V$ and $h_W$ 
\begin{align}
(P_a^{\prime}F_a^{V \prime})h_V^{\prime} 
+ (P_a^{\prime}F_a^{W \prime})h_W^{\prime} 
= 
(P_aF_a^{V})h_V
+ (P_aF_a^{W})h_W . 
\label{PVPW}
\end{align}

By using Eq. (\ref{vector}) for Eq. (\ref{PVPW}), 
we obtain 
\begin{align}
\left(
\begin{array}{c}
P_a^{\prime}F_a^{V \prime}\\
P_a^{\prime}F_a^{W \prime}
\end{array}
\right)
&= 
\left(
\begin{array}{cc}
\cos\psi & -\sin\psi\\
\sin\psi & \cos\psi
\end{array}
\right)  
\left(
\begin{array}{c}
P_aF_a^{V}\\
P_aF_a^{W}
\end{array}
\right) .  
\label{vector2}
\end{align}
This means that 
$(P_aF_a^{V}, P_aF_a^{W})$ is a spin-1 vector. 
This spin-1 property can be shown also by 
straightforward calculations 
of using explicit forms of $P_a$, $F_a^V$ and $F_a^W$. 
From Eq. (\ref{vector2}), one can immediately show that 
the squared magnitude as 
$(P_aF_a^V)^2 + (P_aF_a^W)^2$ remains unchanged by the rotation.

Hence, we use the invariant combination 
\begin{align}
PF^{VW} &\equiv \sqrt{(P_aF_a^V)^2 + (P_aF_a^W)^2} , 
\\
QF^{VW} &\equiv \sqrt{(Q_aF_a^V)^2 + (Q_aF_a^W)^2} ,
\label{vector-invariant}
\end{align}
to define, by the same way for Eq. (\ref{suppression}), 
the suppression factor for the vector modes as 
\begin{align}
\rho^{VW} 
\equiv \frac{\max(\mbox{PQF}^{VW}) - \mbox{PQF}^{VW}}
{\max(\mbox{PQF}^{VW})} , 
\label{suppresion}
\end{align}
where $\mbox{PQF}^{VW}$ denotes that larger one of $PF^{VW}$ and $QF^{VW}$ 
and $\max(\mbox{PQF}^{VW})$ is the largest one of $\mbox{PQF}^{VW}$ 
for the observed events.  
Here, the magnitude of $(P_aF_a^{V}, P_aF_a^{W})$ is corresponding to 
the scalar counterpart as $P_aF_a^S$. 

\begin{figure}
\includegraphics[width=16cm]{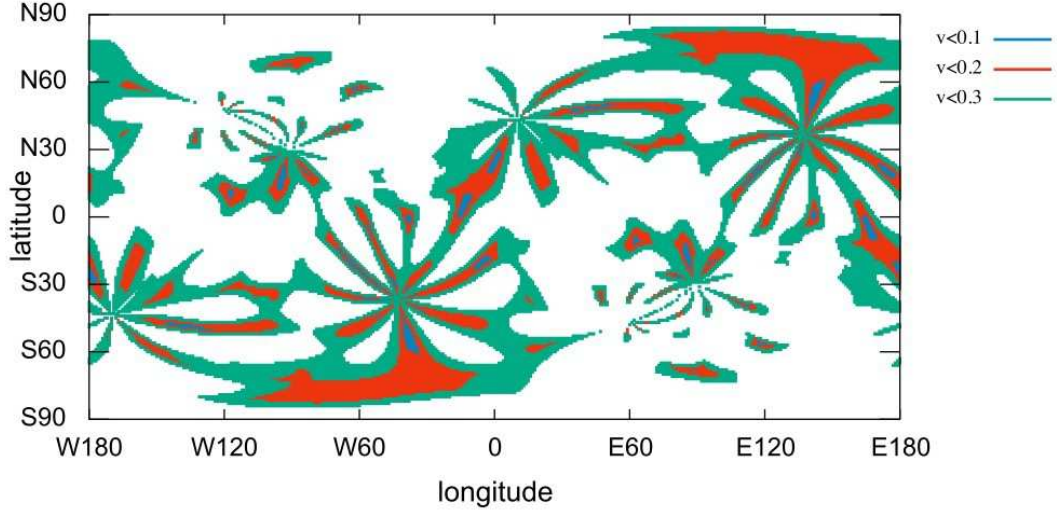}
\caption{
Contour map of 
$\mbox{PQF}^{VW}$, namely 
the larger one between $\sqrt{(P_aF_a^V)^2 + (P_aF_a^W)^2}$ 
and $\sqrt{(Q_aF_a^V)^2 + (Q_aF_a^W)^2}$  
in the null stream 
for the HLVK network. 
The vertical and horizontal axes denote 
the latitude and longitude of a GW source, respectively. 
The maximum value of each of the vector coefficients is unity. 
}
\label{figure-contour-vector}
\end{figure}

Figure \ref{figure-contour-vector} is the contour map for 
$\sqrt{(P_aF_a^V)^2 + (P_aF_a^W)^2}$. 
This figure shows that 
GW events with the suppression factor 
$\rho^{VW} \geq 0.9$ 
occur 
in nearly one percent of the whole sky area. 
This figure implies that 
the suppression of the vector modes 
occurs less frequently than that of the scalar modes. 
For instance,  
the area of the suppression of the vector modes as 
$\rho^{VW} \geq 0.9$ 
is much smaller than that for the scalar modes 
as $\rho^{S} \geq 0.9$. 
This contrast comes from the difference in the number of 
the related coefficients in the null streams: 
The coefficient for the scalar is $P_aF_a^S$ in Eq. (\ref{PS}), 
while those for the vector modes are $P_aF_a^V$ and $P_a F_a^W$.  
As a result, 
the probability that both $P_aF_a^V$ and $P_aF_a^W$ are simultaneously 
small enough to achieve a large $\rho^{VW}$ is small. 
 
\begin{figure}
\includegraphics[width=16cm]{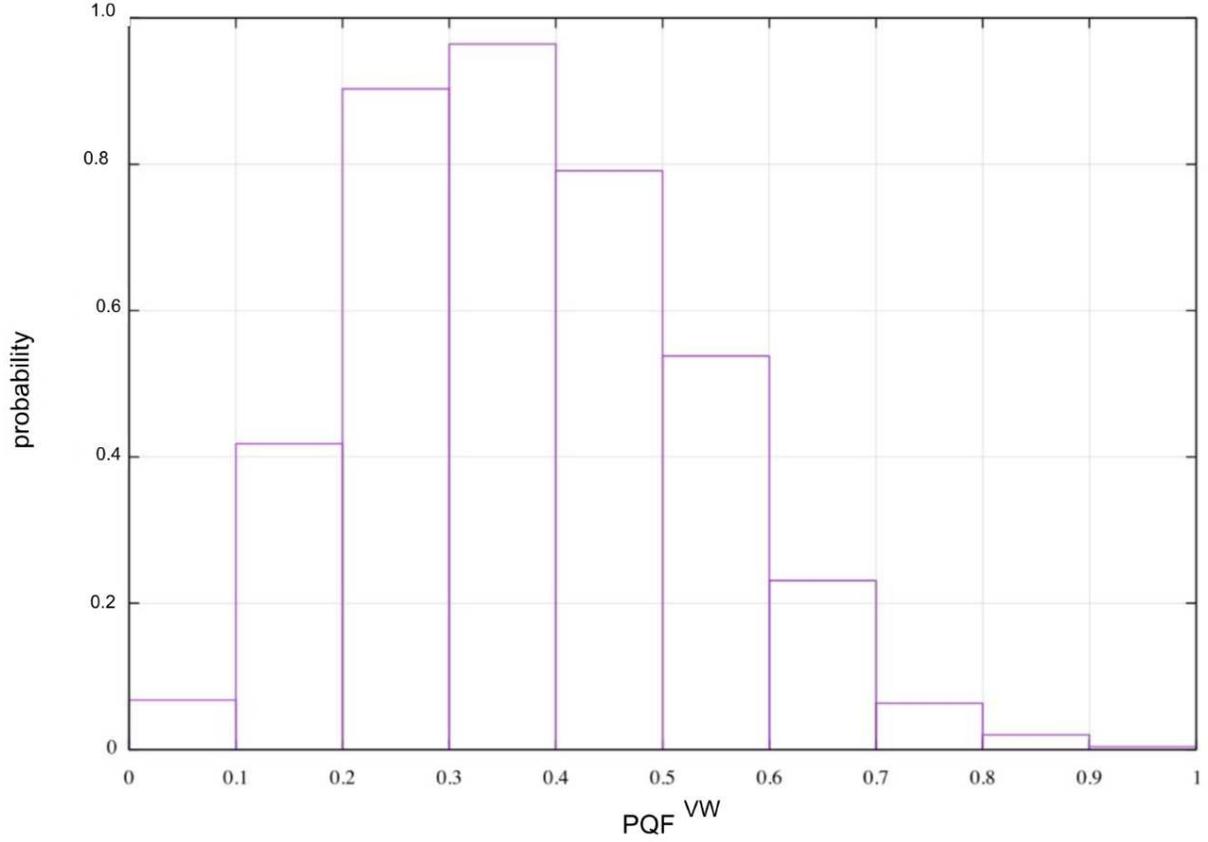}
\caption{
The vertical axis denotes the probability  
for each bin, 
where the horizontal axis denotes the largest vector coefficient 
in the null streams. 
We prepare numerically a random distribution of 10,000 events 
in the sky. 
The probability  
distribution has a single peak around $\sim 0.3$. 
The overall behavior of this histogram is confirmed numerically 
by changing the total number of events such as 20,000 and 100,000. 
}
\label{figure-histogram-vector}
\end{figure}

\begin{figure}
\includegraphics[width=16cm]{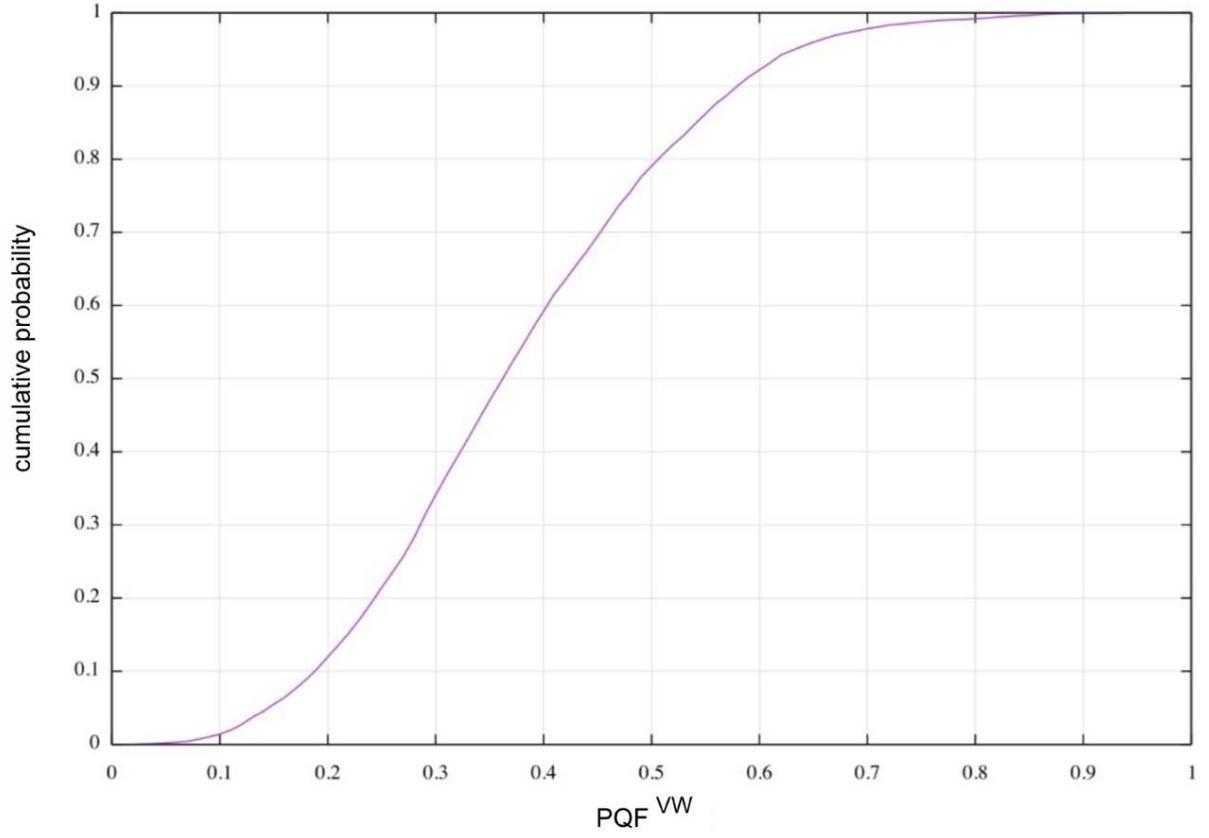}
\caption{
Event probability  
v.s. the threshold of the suppression factor for the vector mode. 
The cumulative probability 
in the vertical axis increases 
as the threshold for the suppression factor is larger. 
}
\label{figure-accumulate-vector}
\end{figure}

Figure \ref{figure-histogram-vector} is 
a histogram of GW events
for a random distribution of GW sources in the sky, 
where the horizontal axis denotes 
the larger one of $\sqrt{(P_aF_a^V)^2 + (P_aF_a^W)^2}$ 
and $\sqrt{(Q_aF_a^V)^2 + (Q_aF_a^W)^2}$. 
In Figure \ref{figure-histogram-vector} , there is an excess around 
$0.2 \sim 0.5$. 
Therefore, we can expect that the probability of testing the spin-0 modes 
with suppressing the vector modes is not low. 
We plot also the 
cumulative probability distribution of events.  
See figure \ref{figure-accumulate-vector}. 
If we choose the threshold for the suppression factor 
$\rho^{VW}$ on the vector modes as 
$\rho^{VW} \geq 0.8$ (corresponding to $\mbox{PQF}^{VW} \leq 0.2$)  
for instance, then,  
nearly one of ten events can be used for a practical test 
of 
spin-0 
polarizations. 
If we choose a tighter threshold as 
$\rho^{VW} \geq 0.9$  
(corresponding to $\mbox{PQF}^{VW} \leq 0.1$), 
only 2 percents of the events can be used 
for a separate test of spin-0 modes. 
Roughly speaking, 
the event rate for the suppression of vector modes 
is smaller by a factor of nearly five than that of the scalar suppression, 
as shown by Figures $\ref{figure-accumulate-scalar}$ and 
$\ref{figure-accumulate-vector}$.

\section{GW170817}
In this section, we mention GW170817 event \cite{GW170817}. 
For this event, 
HKV made observations, while KAGRA was under construction. 
Hence, the null stream for $P_a$ can be used for real data analysis 
of GW170817, 
while that for $Q_a$ will be used only for a theoretical interest 
but not for any real data analysis. 
The coefficients in the null streams are 
estimated as 
$P_a F_a^S = -0.0738$, 
$P_a F_a^V = 0.3772$, 
$P_a F_a^W = 0.3245$, 
$Q_a F_a^S = 0.0091$, 
$Q_a F_a^V = 0.2924$, 
$Q_a F_a^W = 0.3768$, 
where the reference for polarization angles is chosen 
as aLIGO-Livingston ($a=2$). 
Therefore, 
$\sqrt{(P_aF_a^V)^2 + (P_aF_a^W)^2} = 0.4976$, 
and 
$\sqrt{(Q_aF_a^V)^2 + (Q_aF_a^W)^2} = 0.4769$. 
Interestingly, the sky position of GW170817 is 
very particular, in the sense that the scalar coefficient 
$P_a F_a^S$ for the HLV network is significantly small. 
If $|P_a|$ is small, however, 
not only $P_a F_a^S$ but also $P_a F_a^V$ and $P_a F_a^W$ are small. 
For such a case, one cannot test separately the scalar and vector modes. 
If the coefficients in the null streams are normalized 
by the magnitude of $P_a$ or $Q_a$ in order to exclude 
the case of small $|P_a|$ or $|Q_a|$, 
they are 
$P_a F_a^S/|P_a| = -0.1540$, 
$P_a F_a^V/|P_a| = 0.7868$, 
$P_a F_a^W/|P_a| = 0.6768$, 
$Q_a F_a^S/|Q_a| = 0.0212$, 
$Q_a F_a^V/|Q_a| = 0.6853$, 
$Q_a F_a^W/|Q_a| = 0.8831$, 
$\sqrt{(P_aF_a^V)^2 + (P_aF_a^W)^2}/|P_a| = 1.037$, 
and 
$\sqrt{(Q_aF_a^V)^2 + (Q_aF_a^W)^2}/|Q_a| = 1.118$. 
The scalar coefficients are thus much smaller than the vector ones, 
even after they are normalized.

\begin{figure}
\includegraphics[width=12cm]{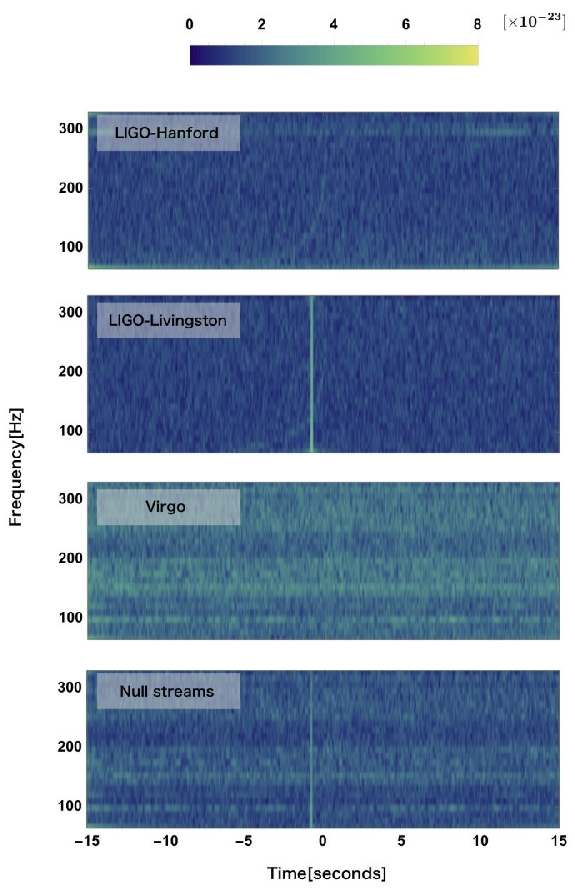}
\caption{
Time-frequency representations 
for the GW170817 \cite{GW170817-data}: 
From top to bottom, 
aLIGO-Hanford, aLIGO-Livingston, and Virgo 
and the null stream $|P_aS_a|$ for the HLV network,  
where the arrival time at 
aLIGO-Hanford (and Virgo) 
is shifted with respect to 
aLIGO-Livingston. 
Times are shown relative to August 17, 2017 12:41:04 UTC.
The vertical line (yellow green in color) around 
$-0.7$ seconds is due to 
the glitch which is a brief burst of instrumental noise a few seconds 
prior to event peak at aLIGO-Livingston, 
where the coalescence time is at time 0.4 seconds in this figure 
\cite{GW170817}. 
A horizontal striped pattern (light green in color) is 
due to noises at each detector. 
No chirp-like events are seen 
above the level of $\sim 2 \times 10^{-23}$ in the null streams. 
}
\label{figure-nullstream}
\end{figure}

Figure \ref{figure-nullstream} shows the null stream 
for $P_a$ of the three detector outputs 
\cite{GW170817-data} 
for GW170817. 
We examine the GW data in the time domain from $-15$ to $+15$ seconds 
around the GW170817 event, for which 
the velocity of the vector GWs (denoted as $c_g^{\mbox{vect}}$) 
is limited within 
\begin{align}
\left|\frac{c_g^{\mbox{vect}} - c}{c}\right| 
\leq 4 \times 10^{-15} 
\left(\frac{40 \mbox{Mpc}}{D}\right) 
\left(\frac{|\delta t|}{15 \mbox{sec.}}\right) , 
\end{align}
where $D$ is the distance to the GW source and 
$\delta t$ is the arrival time difference between the 
spin-2 
and spin-1 waves. 
In this figure, there are no chirp-like signals 
around the GW arrival time. 
This is consistent with that spin-2 modes cancel out 
in the null stream. 
Roughly speaking, $|P_aS_a| < 2 \times 10^{-23}$ 
in Figure \ref{figure-nullstream}. 
Therefore, an upper bound on the vector GWs can be placed as 
$|h^V + h^W| < 6 \times 10^{-23}$, 
where we use an approximation as $P_a F_a^V \sim P_a F_a^W \sim 1/3$ 
in Eq. (\ref{PS}). 
If the GW detector network including KAGRA observes 
a GW170817-like event with an EM counterpart in the future, 
the use of two null streams will be able to put a tighter constraint on 
the vector modes separately. 

On the other hand, the vector coefficients in the null stream for GW170817 
are not so small. 
Therefore, a non-trivial constraint on 
the scalar modes is not placed by this event.  
It is trivial that the scalar modes must be smaller than 
the tensor modes.

\section{Summary} 
In hope of the near-future network of Advanced LIGO, 
Advanced Virgo and KAGRA, 
we discussed a possibility that the spin-0 modes are 
substantially suppressed in the null streams 
and thus the spin-1 modes can be tested within the noise level. 
We studied also a possibility that the spin-1 modes are  
suppressed in the null streams and thus the spin-0 modes become separately testable. Our numerical calculations show that,  
for one of five events, 
the scalar parts in the null streams are suppressed 
by a factor of ten or more, so that such a suppressed event 
can be used for a test of the spin-1 modes separately from the other modes. 
On the other hand, the possibility of testing the spin-0 modes separately 
from the other modes seems much lower. 
For nearly one of twenty events, the spin-1 modes are significantly suppressed 
by a factor of five or more and thus can be used for a test of the scalar GWs. 
The scalar coefficient in the null stream for HLV observations 
of GW170817 is so small that an upper bound on the amplitude of the vector GWs can 
be put as $< 6 \times 10^{-23}$. 
It will be left for future work to put a more severe constraint 
on the vector modes, if the future GW detector network including KAGRA 
observes a GW170817-like event with an EM counterpart.

\begin{acknowledgments}
We wish to thank Seiji Kawamura, Nobuyuki Kanda and Hideyuki Tagoshi, Yousuke Itoh 
and Yasusada Nambu for fruitful discussions. 
We would like to thank 
Yuuiti Sendouda, Yuya Nakamura and Ryunosuke Kotaki 
for the useful conversations. 
This research has made use of data, software and/or web tools obtained from the Gravitational Wave Open Science Center (https://www.gw-openscience.org), a service of LIGO Laboratory, the LIGO Scientific Collaboration and the Virgo Collaboration. LIGO is funded by the U.S. National Science Foundation. Virgo is funded by the French Centre National de Recherche Scientifique (CNRS), the Italian Istituto Nazionale della Fisica Nucleare (INFN) and the Dutch Nikhef, with contributions by Polish and Hungarian institutes. 
A.N. is supported by JSPS KAKENHI Grant Nos. JP17H06358 and JP18H04581. 
This work was supported 
in part by Japan Society for the Promotion of Science (JSPS) 
Grant-in-Aid for Scientific Research, 
No. 17K05431 (H.A.), 
and 
in part by Ministry of Education, Culture, Sports, Science, and Technology,  
No. 17H06359 (H.A.).  
\end{acknowledgments}

\newpage

\newpage

\newpage

\newpage

\newpage

\newpage

\newpage

\newpage
\begin{figure}
\includegraphics[width=16cm]{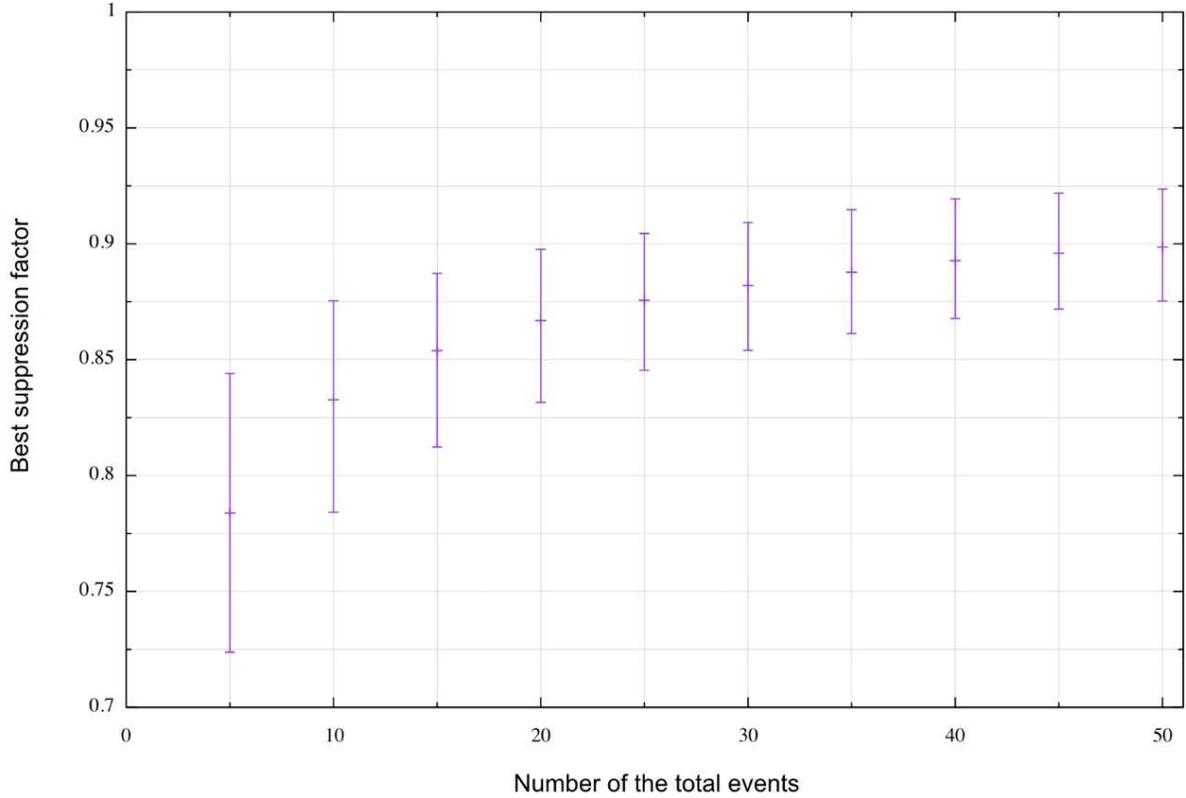}
\caption{
The vertical axis denotes the best suppression factor for the vector modes, 
while the horizontal axis denotes the total event number. 
The error bars represent $1\sigma$ deviation. 
For a few events, for instance, 
the best suppression factor is expected to be around $\sim 0.8$.  
If the number of observed events are increased up to ten for example, 
the best suppression may be $\sim 0.8$ or more, 
which means that a contribution of 
the vector modes in the null streams is suppressed 
by a factor of nearly five. 
Such an event may be used marginally 
for a separate test of the spin-0 modes. 
}
\label{figure-eventnumber-vector}
\end{figure}

\newpage


\begin{thebibliography}{99}
\bibitem{footnote2}
This is true for current ground-based detectors. 
However, it is not true for, e.g. planned Einstein Telescope or LISA, 
where the arms are not at 90 degrees with respect to each other.
\end{thebibliography}
\end{document}